\documentclass[twocolumn,showpacs,preprintnumbers,amsmath,amssymb]{revtex4}
\usepackage{graphicx}% Include figure files
\usepackage{dcolumn}% Align table columns on decimal point
\usepackage{bm}% bold math

\begin{document}

\preprint{APS/123-QED}

\title{A swollen phase observed between the liquid-crystalline phase and the interdigitated phase induced by pressure and/or adding ethanol in DPPC aqueous solution}% Force line breaks with \\

\author{Hideki Seto}
\email{st@scphys.kyoto-u.ac.jp}
 \affiliation{Department of Physics, Kyoto University, Kitashirakawa, Sakyo, Kyoto 606-8502, Japan}%Lines break automatically or can be forced with \\
 
\author{Hidekazu Nobutou}%
\altaffiliation[Present address: ]{Hiroshima Elpida Memory Inc., Higashihiroshima, 739-0153 Japan}
\affiliation{%
Graduate School of Bio-Sphere Science, Hiroshima University, 1-7-1 Kagamiyama, Higashihiroshima  739-8521, Japan
}%
\author{Mafumi Hishida}
 \affiliation{Department of Physics, Kyoto University, Kitashirakawa, Sakyo, Kyoto 606-8502, Japan}
 
 \author{Norifumi L. Yamada}
\affiliation{
Institute of Materials Structure Science, High Energy Accelerator Research Organization, 1-1 Oho, Tsukuba, 305-0801, Japan
}%
\author{Michihiro Nagao}%
\affiliation{%
Institute for Solid State Physics, The University of Tokyo, 106-1 Shirakata, Tokai 319-1106, Japan
}%
\author{Takayoshi Takeda}%
\affiliation{%
Faculty of Integrated Arts and Sciences, Hiroshima University, 1-7-1 Kagamiyama, Higashihiroshima 739-8521, Japan
}%

date: {\today}% It is always \today, today,
\pacs{61.12.Ex, 64.70.Nd, 87.15.-v, 87.14.Cc, 87.16.Dg}% PACS, the Physics and Astronomy
\begin{abstract}
A swollen phase, in which the mean repeat distance of lipid bilayers is larger than the other phases, is found between the liquid-crystalline phase and the interdigitated gel phase in DPPC aqueous solution. Temperature, pressure and ethanol concentration dependences of the structure were investigated by small-angle neutron scattering, and a bending rigidity of lipid bilayers was by neutron spin echo. The nature of the swollen phase is similar to the anomalous swelling reported previously. However, the temperature dependence of the mean repeat distance and the bending rigidity of lipid bilayers are different. This phase could be a precursor to the interdigitated gel phase induced by pressure and/or adding ethanol.
\end{abstract}

\maketitle

\clearpage
\section{\label{sec:intro}Introduction}
Lipid bilayers are interested in these decades as model biological membranes not only from a viewpoint of biology but also of physics. \cite{Nagle04} They exhibit a richness of structures and phase equilibria depending on their environmental conditions such as water content, ionic strength, temperature, pressure, etc. A fluid lamellar phase (liquid-crystalline $L_{\alpha}$ phase) is a basic structure of biological membranes appears at higher temperature. In this phase, bilayers are regularly stacked and flat on average, forming multi-lamellar vesicles. With decreasing temperature, several thermotropic phase transitions have been observed; a "main transition" from the liquid-crystalline phase to a gel ($P_{\beta}^{\prime}$) phase and a "pre-transition" from the $P_{\beta}^{\prime}$ to another gel ($L_{\beta}^{\prime}$) phase. In these gel phases, the hydrophobic tails of lipid molecules are extended and ordered, whereas the tails are conformationally disordered in the liquid-crystalline phase. In the middle-temperature $P_{\beta}^{\prime}$ phase, a two-dimensional lattice structure is formed in which the lipid bilayers are distorted by a periodic ripple in the plane of lamellae.

Another interesting structural polymorphism is an interdigitated gel ($L_{\beta I}$) phase, which is characterized by an alternative alignment of hydrophilic headgroups and hydrophobic tails of phospholipid molecules.  Addition of small molecules, for example alcohol, is known to stabilize the interdigitated structure. \cite{Simon84} An origin of the stabilization by adding alcohol was explained with  the experimental results by Adachi ${et \ al.}$, \cite{Adachi95} which indicated that two alcohol molecules are filled in a volume surrounded interstitially by the headgroups of phospholipid molecules. This picture was verified by a simulation done by Kranenburg and Smit. \cite{Kranenburg04}

The effects of pressure on lipid membranes have been interested in relation to the pressure-anesthetic antagonism; pressure may suppress the effect of anesthesia molecules on structures of bio-membranes. From this viewpoint, Kaneshina ${et \ al.}$ investigated the effect of ethanol on the $P$-$T$ phase diagram of lipids including dipalmitoyl-sn-glycero-3-phosphatidylcholine (DPPC) aqueous solution. \cite{Kaneshina98} They have shown that the interdigitated phase appeared at lower pressure with adding small amount of ethanol. This result implied that the effects of pressure and ethanol on the structural formation of DPPC membranes were the same. However, the detailed mechanisms of the ethanol-induced and the pressure-induced interdigitation and their equality remain to be answered. \cite{Winter01}

In order to compare the effects of adding ethanol and increasing pressure on the structural formation of lipid membranes, comprehensive set of small-angle neutron scattering (SANS) data on the temperature, pressure and ethanol concentration dependences of the structure of DPPC aqueous solution were collected. The results indicated that the phase boundary of the liquid crystalline phase (${L_{\alpha}}$) and the interdigitated phase (${L_{\beta I}}$) was not changed by adding ethanol. However, the transition pressure from the gel phases (${L_{\beta}^{\prime}}$ and  ${P_{\beta}^{\prime}}$)  decreased by adding ethanol. 

A new ``swollen phase", in which the thickness of lipid bilayers is almost the same as in the gel phases and only the thickness of the water layer is larger than the other phases, is found between the ${L_{\alpha}}$ phase and the  ${L_{\beta I}}$ phases. \cite{Seto03} This phase is induced only by increasing pressure (without adding ethanol) and only by adding ethanol (without increasing pressure). The nature of this phase is similar to the anomalous swelling observed between the liquid crystalline phase and the gel phase;  \cite{Chu05} however, the bending rigidity estimated by neutron spin echo (NSE) experiments indicated that the bilayer is not soften unlike the case of the anomalous swelling.

%%%%%%%%%%%%%%%%%%%%%%%%%%%%%%%%%%%%%%%%%%
\begin{figure}
\includegraphics[width=0.35\textwidth]{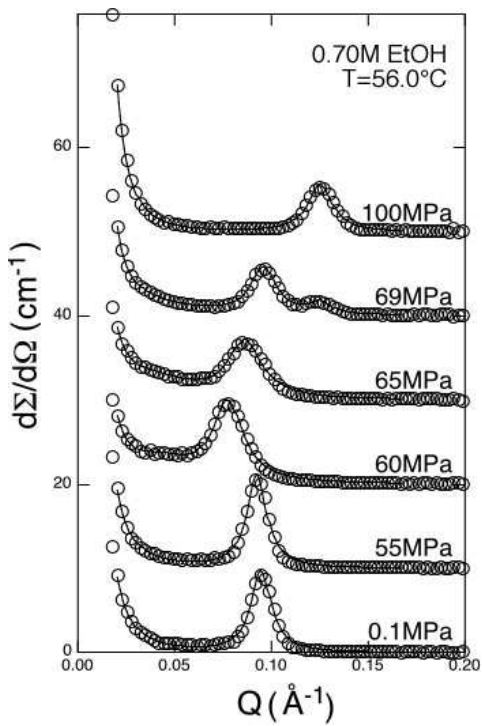}% Here is how to import EPS art [height=.25\textheight]
\caption{\label{fig:1} 
Pressure-dependence of the SANS profile observed at $T = 56.0 ^{\circ}\mathrm{C}$ for the sample with 0.70 M ethanol. The vertical axis indicates the absolute scattering intensity, and shifts 10 cm$^{-1}$ each for better visualization. The liquid crystalline phase transforms to the ``swollen phase" at $P = 60 \mathrm{MPa}$, and to the interdigitated phase above $P = 69 \mathrm{MPa}$. 
}
\end{figure}
%%%%%%%%%%%%%%%%%%%%%%%%%%%%%%%%%%%%%%%%%%

\section{\label{sec:exp}Experiment}

DPPC was purchased from Avanti Polar Lipids, Inc. Deuterated water of 99.9 \% purity and ethanol (EtOH) of 99 \% were obtained from Isotec Inc. and Katayama Chemical Co. Ltd., respectively. All the chemicals were used without further purification. DPPC was first milled in an agate mortar and dissolved in water at 15 wt.\%. Molar ratios of ethanol against water were 0.00 M, 0.30 M, 0.50 M, 0.70 M, 0.80 M, and 1.20 M for SANS and 1.21M for NSE. All the mixtures were incubated at 50 $^{\circ}$C (above the main transition temperature) for 24 hours and kept at room temperature for more than 24 hours.

The SANS experiments were carried out at SANS-U 
installed at the cold neutron guide hall of JRR-3M in JAEA (Japan Atomic Energy Agency), Tokai, Japan. \cite{ItoPB, Okabe}
A 7.0 \AA\ beam of incident cold neutrons with the wavelength resolution of about 10\% 
was used. The camera length for the samples without ethanol and with 0.30 M, 0.50 M and 0.70 M ethanol was 1.5m, that for the sample with 0.80 M ethanol was 1 m, and that for the sample with 1.20 M ethanol was 2 m. The measured  momentum transfer $Q$ ranges for those conditions were  $0.017\le Q\le0.17$ \AA$^{-1}$,  $0.03\le Q\le0.3$ \AA$^{-1}$, and  $0.014\le Q\le0.14$ \AA$^{-1}$, respectively. Exposure time for one measurement was between 300 and 1800 s. 

The samples were contained in two types of high-pressure cells. The first one, which achieves up to 100 MPa, \cite{Takeno, KawabataPRL} was used for the samples with higher ethanol content (0.70 M, 0.80 M, and 1.20 M). The other one, which could be pressurized over 200 MPa, was used for the samples with low ethanol content because the transition pressure to the interdigitated phase increases with decreasing ethanol. In all the cases, the sample thickness was 1 mm.
Hydrostatic pressure was applied by using a hand pump 
and the pressure was measured with a HEISE gauge within an accuracy of $\pm 0.1$ MPa. Temperature of the high-pressure cells was controlled with a personal computer and its accuracy was less than $\pm 0.01\ ^{\circ}$C.
The observed 2-dimensional scattering intensities were azimuthally averaged, 
corrected for the transmission, background scattering and sample thickness, 
and were scaled to the absolute differential scattering cross sections 
by using a secondary standard sample made of Lupolen 
(a polyethylene slab calibrated with the incoherent scattering intensity of Vanadium).

The NSE measurements were performed at iNSE in JAEA, Tokai, Japan. \cite{Nagao06} A 7.1 \AA \ incident neutron was mechanically selected by a neutron velocity selector with the resolution of 12 \%, so that the ranges of momentum transfer and Fourier time are $0.05\le Q\le0.13$ \AA$^{-1}$ and 0.08 $\le t \le 15$ ns, respectively. The sample with 1.21 M ethanol was used and measured at 36.0, 38.5 and 41.0 $^{\circ}$C controlled by a water circulation system within the accuracy of 0.1 $^{\circ}$C.

\section{\label{sec:result}Results and Discussion}
In Fig. 1, a typical example of a pressure-dependence of SANS profile  is shown. In this measurement, 0.70 M of ethanol was added in the DPPC aqueous solution and temperature was kept at 56.0 $^{\circ}$C. The peak positions of these profiles are obtained by the fitting with the function proposed by Mason ${et \ al.}$, \cite{Mason99}

\begin{equation}
    I_{\mathrm{Mason}} = I_{0}Q^B + \frac{I_l}{\xi_l^2(Q-Q_0)^2+1}
\end{equation}\label{eq_Mason}

\noindent where the first term means a diffuse scattering around $Q=0$ originated from the concentration fluctuation of lipids, and the second term represents the Bragg peak due to the lamellar structure.

The phase at ambient pressure is the liquid-crystalline $L_{\alpha}$ and the SANS profile has a Bragg peak at $Q = 0.095 \mathrm{\AA^{-1}}$ due to the regular stacking of lipid bilayers. This profile does not change up to $P=55$ MPa. Above $P=60 \mathrm{MPa}$, the peak at $Q=0.095\ \mathrm{\AA^{-1}}$ disappears and a new peak is born at $Q=0.078\ \mathrm{\AA^{-1}}$. This peak shifts to higher-$Q$ with increasing pressure, coexists with the peak of the $L_{\beta I}$ phase at $P=69\ \mathrm{MPa}$, and vanished above $P=70\ \mathrm{MPa}$. As described in our previous paper, \cite{Seto03} this profile is typical for the new phase existed between the liquid-crystalline phase and the interdigitated gel phase. At high pressure above 70 MPa, a Bragg peak exists at $Q = 0.125  \mathrm{\AA^{-1}}$ because the interdigitated gel phase ($L_{\beta I}$) is induced. 

In order to characterize these structures, further fittings with the scattering function given by Lemmich ${et \ al.}$ \cite{Lemmich96} were performed. They described the scattering function from bilayer structure as,

\begin{equation}
I_{\mathrm{Lemmich}}(Q)\propto\frac{1}{Q^4}
\left\{\displaystyle i_B(Q)+\frac{i_C(Q)}{N}\right\}
\end{equation}\label{eq_Lemmich}

\noindent where

\begin{widetext}
\begin{eqnarray}
i_B(Q)=\Re\left[\frac{(1-F_W)(1-F_H^2F_L)+(\rho_r-1)^2(1-F_L)(1-F_H^2F_W)}{1-F_D}\right] \qquad \nonumber \\
+\Re\left[\frac{2(\rho_r-1)F_H(1-F_W)(1-F_L)}{1-F_D}\right] \\
i_C(Q)=\Re\left[F_W(1-F_D^N)\left(\frac{(1-F_H^2F_L)+(\rho_r-1)F_H(1-F_L)}{1-F_D}\right)^2\right] \qquad
\end{eqnarray} 

\noindent and $F_{H}$, $F_{L}$, $F_{W}$, $F_{D}$ are

\begin{equation}
 F_\nu(Q)=\exp\left[-iQd_\nu-\frac{Q^2\sigma_\nu^2}{2}\right](\nu=H,L,W,D),
\end{equation}
\end{widetext}

\noindent with $d$ and $\sigma$ being thickness and standard deviation. The suffixes ${H, L, W}$ are of headgroup, hydrocarbon chain and water, and ${D}$ means a whole bilayer, i.e., the mean repeat distance of lamellar $d$ is equal to $d_D=2d_H+d_L+d_W$. Here $\rho_{r}$ is described as,

\begin{equation}
\rho_{r}=(\rho_{L}-\rho_{W})/(\rho_{H}-\rho_{W})
\end{equation}

\noindent with the scattering amplitude densities of these parts. 

The fit function explains the profiles at the single phases well. The estimated thickness $t$'s of a lipid bilayer were 43.4 \AA  \ for $L_{\alpha}$ phase, 32.7 \AA \ for $L_{\beta I}$ phase, 47.3 \AA \ for $P_{\beta}^{\prime}$ phase, and 47.1 \AA \ for $L_{\beta}^{\prime}$ phase, respectively. These values are consistent with those in the literature. \cite{Adachi95, Inoko78, Braganza86}

The thickness of a bilayer in the new phase is estimated to be 47.5 \AA, which is almost the same as that in the gel phases ($P_{\beta}^{\prime}$ and $L_{\beta}^{\prime}$) . Therefore, the intra-layer packing of lipid molecules in this new phase should be the same as that in the gel phases, and only the thickness of  water layer between bilayers, 33.0 \AA \ at $P=60\ \mathrm{MPa}$, is larger than the other phases. Therefore, here we named this phase as a ``swollen phase" ($L_{s}$ phase).

%%%%%%%%%%%%%%%%%%%%%%%%%%%%%%%%%%%%%%%%%%
\begin{figure*}
\includegraphics[width=0.9\textwidth]{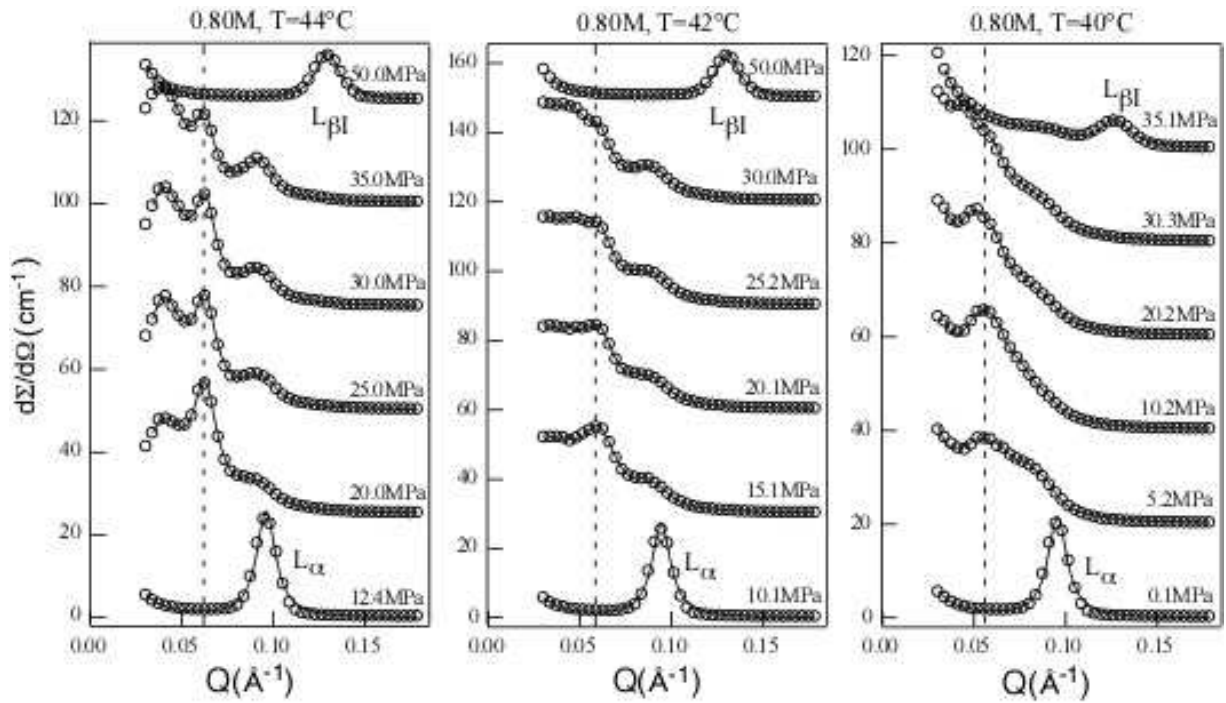}% Here is how to import EPS art [height=.25\textheight]
\caption{\label{fig:2} 
Pressure- and temperature dependences of the SANS profile for the sample with 0.80 M ethanol. Each peaks at higher pressure is shifted upward for better visualization. The peak due to the swollen phase ($Q \simeq 0.06$ \AA$^{-1}$) appeared between $T = 40 ^{\circ}\mathrm{C}$ and $44 ^{\circ}\mathrm{C} $ with a peak belonging to the $P_{\beta}^{\prime}$ phase ($Q \simeq 0.09$ \AA$^{-1}$). Thus, the swollen phase exists as a coexistence phase with the $P_{\beta}^{\prime}$ phase.
}
\end{figure*}
%%%%%%%%%%%%%%%%%%%%%%%%%%%%%%%%%%%%%%%%%%

In Fig. 2, pressure-dependences of SANS profile of the sample with 0.80 M ethanol at $T = 44  ^{\circ}\mathrm{C}$ , $42  ^{\circ}\mathrm{C}$ , and $40  ^{\circ}\mathrm{C} $ are shown. In these conditions, the swollen phase exists as a coexistence phase with the ripple gel phase. The position of the Bragg peak depends slightly on temperature, on the other hand, changes very little with increasing pressure. 

%%%%%%%%%%%%%%%%%%%%%%%%%%%%%%%%%%%%%%%%%%%
\begin{figure}
\includegraphics[width=0.45\textwidth]{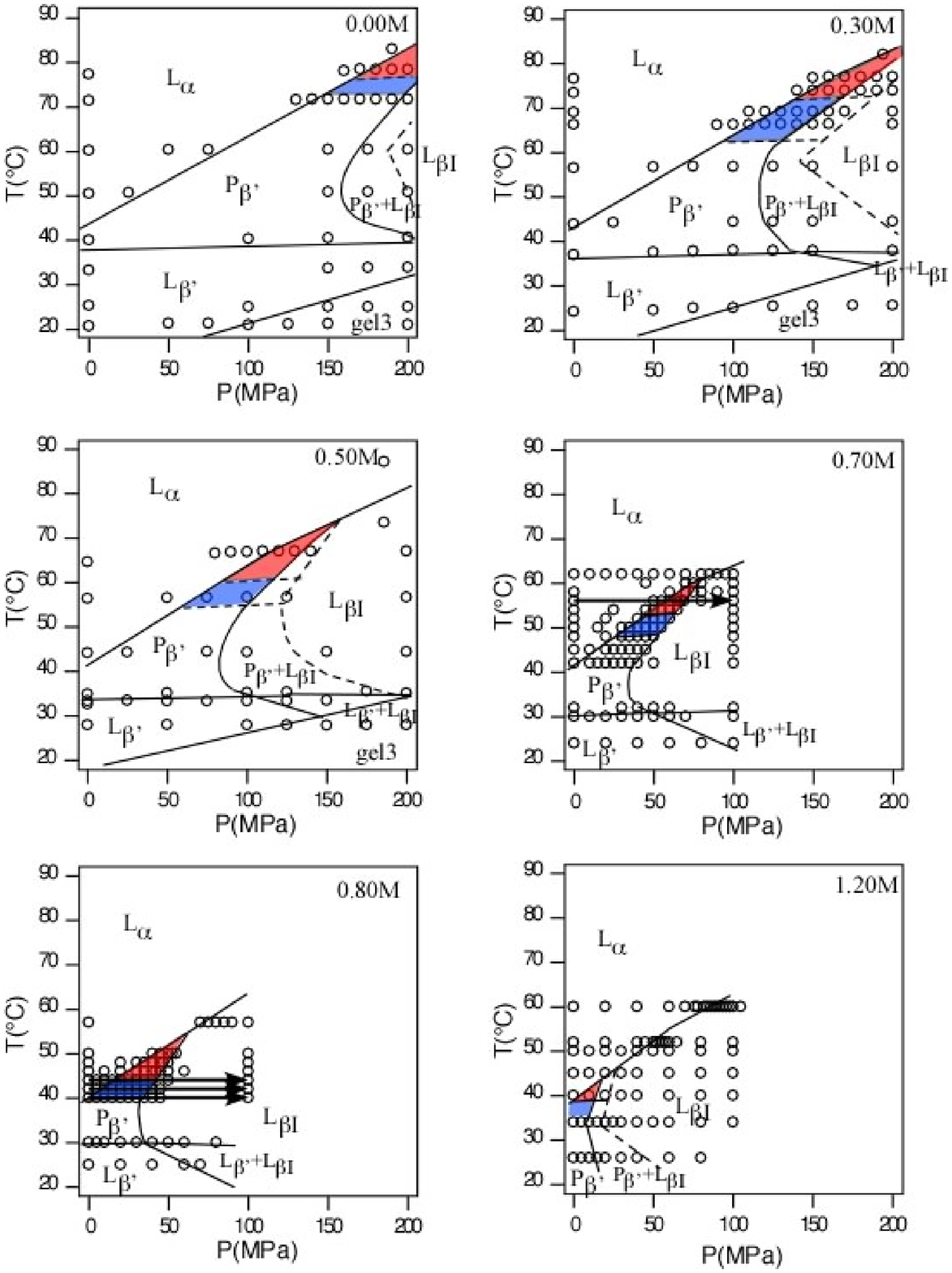}% Here is how to import EPS art [height=.25\textheight]
\caption{\label{fig:3} 
$P-T$ phase diagrams obtained from the SANS experiment. The red areas are the regions of the single swollen phase ($L_{s}$), and the blue areas are the coexistence phase ($L_{s}+P_{\beta}^{\prime}$). The circles indicate the points measured by the present SANS experiment. The arrows indicate the lines of experiments of Fig. 1 and Fig. 2.
}
\end{figure}
%%%%%%%%%%%%%%%%%%%%%%%%%%%%%%%%%%%%%%%%%%

Figure 3 summarizes the phase diagrams determined by the SANS experiments. It is clear that the phase boundary between the liquid-crystalline $L_{\alpha}$ phase and the gel phases does not change by adding ethanol. However, the transition pressure to the interdigitated gel ($L_{\beta I}$) phase decreases with increasing ethanol concentration. Thus, it is could be mentioned that the effect of ethanol on the interdigitation is identical with pressure. The swollen phase, indicated by the red areas (single $L_{s}$ phase) and the blue areas (coexistence with the $P_{\beta}^{\prime}$), exists in all the ethanol concentrations. Both temperature and pressure of these areas decrease with increasing ethanol concentration. This tendency is the same as that of the $L_{\beta I}$ phase. Thus we conclude that the swollen $L_{s}$ phase appears as a precursor to the interdigitated  $L_{\beta I}$ phase. 

%In order to check the reproducibility of the $L_{s}$ phase, we compared the dependences of the SANS profile on increasing and decreasing pressure at constant temperature, and on decreasing temperature and with increasing temperature at ambient pressure. The left part of Fig. 4 shows the pressure-variation SANS profile at $T=41^{\circ}\mathrm{C}$ of the sample with 1.20 M ethanol. The profiles of the $L_{s} + P_{\beta}^{\prime}$ phase in the process of increasing pressure and that of decreasing pressure are identical. However, the temperature variation of the SANS profile depends on the direction; the $L_{s}$ phase appears only with decreasing temperature. This tendency is comparable with the fact that the two-dimensional periodic structure of the $P_{\beta}^{\prime}$ phase only appears with increasing temperature from the $L_{\beta}^{\prime}$ phase. \cite{Wack89, Tenchov89} In the case of the decreasing temperature, the stacking fault of the ripple structure occurs, the system stays at a metastable state whose ripples are not alined and its two-dimensional order is disordered. \cite{Takeda95}

%%%%%%%%%%%%%%%%%%%%%%%%%%%%%%%%%%%%%%%%%%
\begin{figure}
\includegraphics[width=0.45\textwidth]{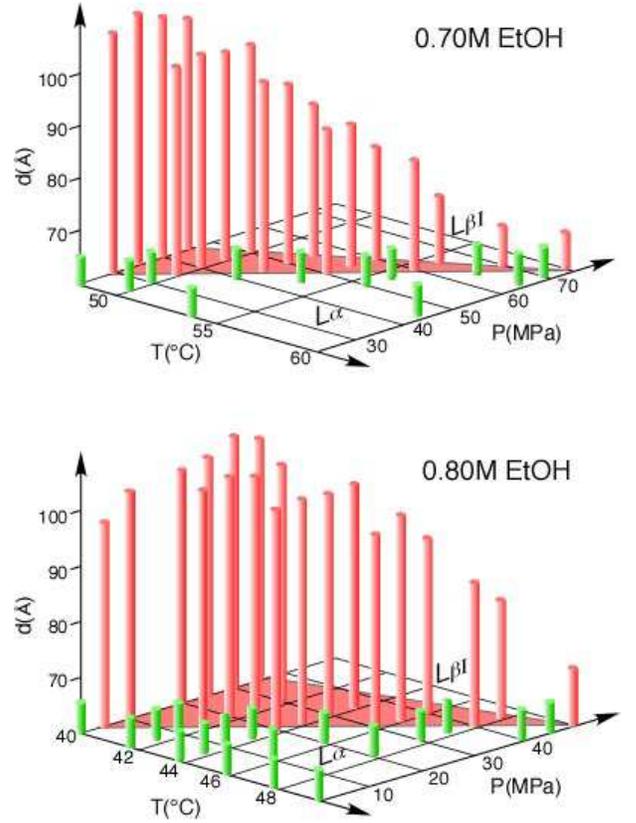}% Here is how to import EPS art [height=.25\textheight]
\caption{\label{fig:4} 
Pressure- and temperature dependences of the mean repeat distance $d$ in the $L_{s}$ phase (red) with that in the $L_{\alpha}$ phase (green) of the samples with 0.70 M ethanol and 0.80 M ethanol. 
}
\end{figure}
%%%%%%%%%%%%%%%%%%%%%%%%%%%%%%%%%%%%%%%%%%

Figure 4 shows pressure- and temperature-dependences of the mean repeat distance $d$ in the $L_{s}$ phase together with $d$ in the $L_{\alpha}$ phase. The pressure-dependence of $d$ is not very significant; it decreases a little with increasing pressure. On the other hand, it decreases monotonically with increasing temperature. This tendency is similar to that of the anomalous swelling observed in similar systems. \cite{Hoenger94, Pabst03, Harroun04} The $d$-spacing of lipid bilayers increases from 63 \AA \ (in the case of DMPC) up to around 67 \AA \ with approaching the main transition temperature from the liquid-crystalline $L_{\alpha}$ phase, and pressure suppresses the anomalous swelling. \cite{Harroun04} However, the details of the temperature dependence of $d$ in the $L_{s}$ phase is different from those of the anomalous swelling: In the case of the anomalous swelling, the mean repeat distance $d$ follows the power law behavior from the $L_{\alpha}$ phase to the main transition temperature with decreasing temperature. On the other hand, in the present case, $d$ jumps at the phase boundary between the $L_{\alpha}$ phase and the $L_{s}$ phase. The value of $d$ in the $L_{s}$ phase is much larger than that in the anomalous swelling. Thus we emphasize that the present observation can not be explained by the anomalous swelling.

Pabst $et \ al.$ indicated that the origin of the anomalous swelling is an abrupt decrease in the bending rigidity of lipid bilayers near the main transition temperature. \cite{Pabst03} Thus, in order to check the rigidity of lipid bilayers of the swollen phase, we investigated dynamical behaviors by means of NSE, because the thermal fluctuation of lipid bilayers can be observed directly by this method.

%%%%%%%%%%%%%%%%%%%%%%%%%%%%%%%%%%%%%%%%%%
\begin{figure*}
\includegraphics[width=0.8\textwidth]{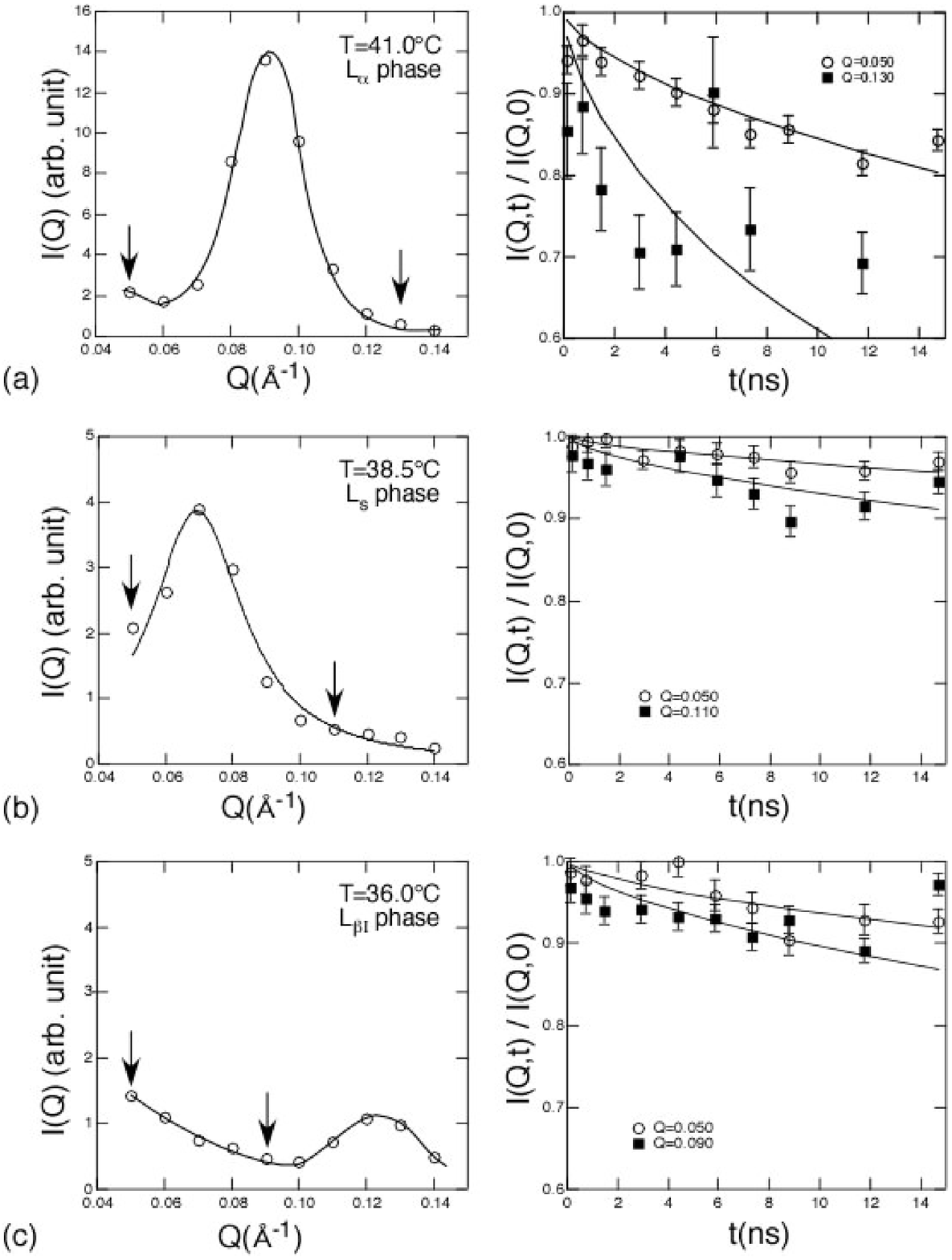}% Here is how to import EPS art [height=.25\textheight]
\caption{\label{fig:5} 
SANS profiles observed at iNSE (left) and the intermediate structure factor $I(Q,t)/I(Q,0)$ (right) obtained from the NSE measurement for the sample with 1.21 M ethanol. (a) $T = 41.0 ^{\circ}$C in $L_{\alpha}$ phase. (b) $T = 38.5 ^{\circ}$C in $L_{s}$ phase. (c) $T = 36.0 ^{\circ}$C in $L_{\beta I}$ phase. The lines in the SANS profile is guides for the eyes, and the lines in the NSE data are the fit results. The arrows in the SANS profiles are the positions where the NSE data are depicted.
}
\end{figure*}
%%%%%%%%%%%%%%%%%%%%%%%%%%%%%%%%%%%%%%%%%%

In Fig. 5, experimental results of the sample with 1.21 M ethanol at ambient pressure obtained by NSE are shown. At $T = 41.0 ^{\circ}$C, the system is in the $L_{\alpha}$ phase and the SANS profile has a Bragg peak at $Q \simeq 0.09$ \AA$^{-1}$. The system transforms to the $L_{s}$ phase at $T = 38.5 ^{\circ}$C in which a peak exists at $Q \simeq 0.07$ \AA$^{-1}$, and to the $L_{\beta I}$ phase at $T = 36.0 ^{\circ}$C with a peak at $Q \simeq 0.12$ \AA$^{-1}$. 

The intermediate structure factors $I(Q,t)/I(Q,0)$ at these temperatures obtained by the NSE experiment are shown in the right part of Fig. 5. In the DPPC aqueous solution, the authors have already showed \cite{YamadaJPSJ05} that $I(Q,t)/I(Q,0)$ can be explained with the theoretical function describing a single membrane fluctuation given by Zilman and Granek, \cite{ZilmanPRL96}

\begin{equation}
I(Q,t)/I(Q,0)=\exp\left[-\left(\Gamma t\right)^{2/3}\right],
\label{eq_lqt_ZG}
\end{equation}

\noindent where $\Gamma$ is proportional to $Q^{3}$

\begin{equation}
\Gamma=A Q^{3}, \label{eq_Gamma}
\end{equation}

\noindent and the bending rigidity of lipid bilayers $\kappa$ is obtained by

\begin{equation}
\kappa=\left(A^{-1}\gamma_{\alpha}\gamma_{\kappa}\left(k_{B}T\right)^{3/2}\eta^{-1}\right)^{2}.
\label{eq_kappaZG}
\end{equation}

\noindent Here, parameters $\gamma_{\kappa}$ and $\gamma_{\alpha}$ could be assumed as 1 and 0.025, respectively. The viscosity of surrounding medium $\eta$ is that of D$_{2}$O. In multi-lamellar systems, a scattering intensity around a Bragg peak will be influenced by a strong elastic scattering, and the nature of the thermal fluctuation of a bilayer should be estimated from off-Bragg region. 

Thus we fitted eq. (\ref{eq_lqt_ZG}) to $I(Q,t)/I(Q,0)$'s of $Q=0.13 \mathrm{\AA}^{-1}$ at $T=41.0^{\circ}\mathrm{C}$, $Q=0.11 \mathrm{\AA}^{-1}$ at $T=38.5^{\circ}\mathrm{C}$, and $Q=0.09 \mathrm{\AA}^{-1}$ at $T=36.0^{\circ}\mathrm{C}$. From these analysis, the coefficient $A$ is estimated to be 15.9 $\mathrm{\AA}^3\mathrm{ns}^{-1}$ at  $T=41.0^{\circ}\mathrm{C}$, 1.44 $\mathrm{\AA}^3\mathrm{ns}^{-1}$ at  $T=38.5^{\circ}\mathrm{C}$, and 4.92 $\mathrm{\AA}^3\mathrm{ns}^{-1}$ at  $T=36.0^{\circ}\mathrm{C}$, respectively. The bending rigidity is estimated from these values of $A$ using eq. (\ref{eq_kappaZG}) to be $\kappa = 1.9 \times 10^{-19}$ J at  $T=41.0^{\circ}\mathrm{C}$, $\kappa = 2.1 \times 10^{-17}$ J at  $T=38.5^{\circ}\mathrm{C}$, and $\kappa = 1.6 \times 10^{-18}$ J at  $T=36.0^{\circ}\mathrm{C}$. The value in the liquid-crystalline phase ($\kappa = 1.9 \times 10^{-19}$ J at  $T=41.0^{\circ}\mathrm{C}$) is consistent with the previous estimation. ($\kappa = 2.56 \times 10^{-19}$ J at  $T=50.0^{\circ}\mathrm{C}$) \cite{YamadaJPSJ05} It is reasonable that the bending rigidity of bilayers in the
interdigitated phase ($\kappa = 2.1 \times 10^{-17}$ J  at  $T=38.5^{\circ}\mathrm{C}$) is larger than that in the gel phases. \cite{KawabataPRL, SetoAIP, footnote} And also, it is clear that the bending rigidity in the swollen phase ($\kappa = 1.6 \times 10^{-18}$ J at  $T=36.0^{\circ}\mathrm{C}$) is harder than that in the liquid-crystalline phase. This is the most striking result in this NSE experiment, that the rigidity of bilayers in the swollen phase is hard as those in the gel phases, unlike the case of the anomalous swelling.

The dynamical behavior of bilayers is also characterized by the fitting to $I(Q,t)/I(Q,0)$'s of $Q=0.05\ \mathrm{\AA}^{-1}$ for all the temperatures observed. The coefficient $A$'s are 55.9 $\mathrm{\AA}^3\mathrm{ns}^{-1}$ at  $T=41.0^{\circ}\mathrm{C}$, 5.0 $\mathrm{\AA}^3\mathrm{ns}^{-1}$ at  $T=38.5^{\circ}\mathrm{C}$, and 13.1 $\mathrm{\AA}^3\mathrm{ns}^{-1}$ at  $T=36.0^{\circ}\mathrm{C}$, respectively. The bending rigidity could not be extracted from these values due to the effect of the elastic scattering; however, the tendency that bilayers in the $L_{s}$ phase is harder than that in the $L_{\alpha}$ phase is clearly verified.

So far, an origin to determine the mean repeat distance of lipid bilayers in multilamellar vesicles is discussed in terms of a balance of interactions between bilayers. Because DPPC is a neutral lipid and no salt is added in the present mixture, the Coulombic interaction should be neglected and only 3 interactions are taken into account; the van der Waals attractive force between two parallel flat bilayers, a short-range repulsive force due to the hydration layer in the vicinity of lipid bilayers, and the long-range repulsive force originated from the thermal fluctuation of lipid bilayers. \cite{Ohshima77, YamadaJPSJ05} If one tries to explain the origin of the swollen phase in this framework, any changes of strengths of interactions comparing with the other phases are necessary. However, there is no reason that the van der Waals force is weaker or the other two repulsive forces are stronger than the other phases. Thus, at present, we can not explain the origin of the swollen phase quantitatively. 

%One hint to answer this question is that the swollen phase does not appear with increasing temperature from the $L_{\beta}^{\prime}$ phase. This tendency is comparable with the meta-stable state of the ripple gel phase, $P_{\beta}^{\prime}$\small{(mst)}. \cite{Takeda95} In the case of the ripple gel phase ($P_{\beta}^{\prime}$), statistically rippled bilayers are in-phase and 2 dimensional periodic structure is stabilized only in the case of increasing temperature. However, when the temperature is decreased from the $L_{\alpha}$ phase, positions and directions of ripples of each layers does not coincide each other and the regular 2 dimensional structure can not be formed spontaneously. Thus, it may be possible that a large static undulation of lipid bilayers due to the irregular ripple structure is an origin of the long-range repulsive force to stabilize the long-period (swollen) structure, because the bilayers in the $L_{s}$ phase is very hard comparing with the thermal undulation.

It should be noted that another long-priod structure named $L_{X}$ phase was observed in the process of the transition from $L_{\alpha}$ phase to $L_{\beta}$ phase in DEPC aqueous solution. \cite{Winter02} This result suggested that a quasi-static potential minimum of the long-spacing structure exists in the vicinity of the $L_{\alpha}$ phase, and the origin of the $L_{X}$ phase could be the same as the present $L_{s}$ phase.

\section{Conclusion}
In this paper, we present SANS results by changing temperature, pressure, and ethanol concentration. It is verified that the adding ethanol has no effect on the phase boundary between the liquid-crystalline phase and the gel phases, on the other hand, have the same effect as pressure on the interdigitation. Between the liquid-crystalline phase and the interdigitated phase, a new swollen phase is observed. The nature of this phase is similar to the anomalous swelling; however, the temperature dependence of the mean repeat distance and the bending rigidity of bilayers are different from those of the anomalous swelling. 

\begin{acknowledgments}
One of the authors (H. S.) acknowledges  Prof. S. Komura, Prof. T. Kato, and Dr. Y. Kawabata at Tokyo Metropolitan University for valuable discussion. The SANS and NSE experiments were done with the approval of the Neutron Scattering Program Advisory Committee (Proposal No. 03.188.f). This work was supported by a Grant-in-Aid for Scientific Research (No. 17540382), by a Grant-in-Aid for the 21st Century COE "Center for Diversity and Universality in Physics" from the Ministry of Education, Culture, Sports, Science, and Technology (MEXT) of Japan, and by the Yamada Science Foundation.
\end{acknowledgments}

\newpage %Just because of unusual number of tables stacked at end


\begin{thebibliography}{99}
\bibitem{footnote}
Normally, pressure reduces the free volume of amphiphilic molecules to affect the packing property, and an entropic contribution of hydrophobic tails to the bending rigidity becomes less. Therefore, the bending rigidity of amphiphilic layers increases with increasing pressure. \cite{KawabataPRL, SetoAIP}
\end{thebibliography}
\end{document}